# Gas-Phase Formation of Glycolonitrile in the Interstellar Medium

Luis Guerrero-Méndez,[a] Anxo Lema-Saavedra,[b] Elena Jiménez,[c,d] Antonio Fernández-Ramos,*[a,b] and Emilio Martínez-Núñez*[a]

Our automated reaction discovery program, AutoMeKin, has been utilized to investigate the formation of glycolonitrile (HOCH$_2$CN) in the gas phase under the low temperatures of the interstellar medium (ISM). The feasibility of a proposed pathway depends on the absence of barriers above the energy of reactants and the availability of the suggested precursors in the ISM. Based on these criteria, several radical-radical reactions and a radical-molecule reaction have been identified as viable formation routes in the ISM. Among the radical-radical reactions, OH+CH$_2$CN appears to be the most relevant, considering the energy of the radicals and its ability to produce glycolonitrile in a single step. However, our analysis reveals that this reaction produces hydrogen isocyanide (HNC) and formaldehyde (CH$_2$O), with rate coefficients ranging from (7.3-11.5)×10$^{-10}$ cm$^3$ molecule$^{-1}$ s$^{-1}$ across the temperature range of 10-150 K. Furthermore, the identification of this remarkably efficient pathway for HNC elimination from glycolonitrile significantly broadens the possibilities for any radical-radical mechanism proposed in our research to be considered as a feasible pathway for the formation of HNC in the ISM. This finding is particularly interesing given the persistently unexplained overabundance of hydrogen isocyanide in the ISM. Among the radical-molecule reactions investigated, the most promising one is OH+CH$_2$CNH, which forms glycolonitrile and atomic hydrogen with rate coefficients in the range (0.3-6.6)×10$^{-10}$ cm$^3$ molecule$^{-1}$ s$^{-1}$ within the 10-150 K temperature range. Our calculations indicate that the formation of both hydrogen isocyanide and glycolonitrile is efficient under the harsh conditions of the ISM.

## 1 Introduction

Glycolonitrile (GLN) is a simple organic molecule that has been proposed as a key precursor to the formation of adenine.[1,2] In 2019, the first detection of glycolonitrile in the interstellar medium (ISM) was reported.[3] The detection of GLN highlights the importance of understanding the formation of this molecule, as it can provide insight into the origins of prebiotic molecules in the universe.

The formation mechanism of organic molecules in the ISM can be explained by several types of reactions. Due to the extremely low temperatures of the interstellar clouds (10-150 K), gas-phase reactions are expected to have barrier heights that approach zero with respect to reactants.[4] Reactions can take place on grain surfaces or through gas-phase chemistry. Although the former can account for the formation of molecules in hot cores that surround protostars and young stars, explaining how molecules in cold, dense sources are formed is still a significant challenge.[4]

Gas-phase reactions in the ISM are expected to occur with submerged barriers,[5,6] which typically requires that at least one of the reactants must be a radical or an ion. Furthermore, the stabilization of the formed products is not favoured at the extremely low densities of the ISM, unless radiative stabilization is competitive.[4]

Several studies have investigated various formation pathways of GLN on ices and/or grains.[7-11] In addition, a few gas-phase reactions have been proposed for the formation of GLN; however, all of these mechanisms exhibit significant reaction barriers.[10] To date, no gas-phase reactions with submerged barriers have been proposed for the formation of GLN. In the year 2001, Woon conducted *ab initio* calculations to show that GLN can be formed from the reaction of CH$_2$O+HNC on ice, despite the fact that it exhibits a barrier height of 46 kcal/mol.[11]

Recent calculations by the same author reveal that the reactions between C$^+$ and HCN embedded in the surface of icy grain mantles can explain the formation of GLN.[7] Early computational studies have also explored isomers of C$_2$H$_3$NO.[12,13]

Danger et al. have shown that GLN can be formed (in competition with methanolamine) from ices containing CH$_2$O, NH$_3$ and CN$^-$ at 40 K.[8] It was also demonstrated, using infrared spectroscopy and mass spectrometry, that the reaction between HCN and CH$_2$O is only feasible in the presence of water.[9]

When using a limited chemical reaction network (CRN) in which GLN can only form on dust grains, the resulting GLN abundances are significantly underestimated.[3] This outcome led the authors of the study to recognize that "key gas phase routes for the formation of this molecule might be missing in our chemical network".[3] Recently, Zhao et al. employed the astrochemical code NAUTILUS and an updated gas-grain CRN to demonstrate that adjusting the cosmic-ray ionization rate is necessary to most accurately replicate the observed abundances of GLN.[10]

This work aims to unravel the barrierless GLN formation pathways through theoretical calculations by examining two distinct types of formation reactions:

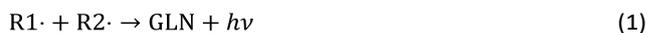

$$R1\cdot + R2\cdot \rightarrow GLN + h\nu \qquad (1)$$

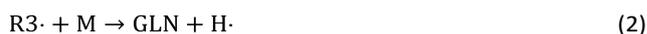

$$R3\cdot + M \rightarrow GLN + H\cdot \qquad (2)$$

Reaction (1), called radiative association, involves two radicals (R1· and R2·) and necessitates effective radiative stabilization of GLN, whereas (2) is a radical-molecule reaction and does not require such stabilization due to the formation of two products: GLN and atomic hydrogen.[4]

To unravel all possible elementary steps involved in the above reactions, we utilize AutoMeKin, an automated program developed in our laboratory for discovering reaction mechanisms in an automated manner.[14] As the decomposition and formation pathways of GLN are interconnected, to study reaction (1) the decomposition pathways of GLN are analyzed. Conversely, investigating the fragmentation mechanisms of HOCH$_2$CHN radical, hereinafter GLN-H, is used to study reaction (2). The decomposition products of GLN (R1· and R2·) and GLN-H (R3· and M) are compared with a database of molecules that have already been detected in the ISM to assess the feasibility of each pathway.

Finally, the most promising mechanisms are evaluated for feasibility by conducting kinetic simulations that replicate the conditions of the ISM.

## 2 Computational details

### 2.1 Automated reaction mechanism mapping

The program AutoMeKin,[14-16] developed in our laboratory, was used to locate the minima and transition states (TSs) on the potential energy surface (PES) of GLN and GLN-H. Our open-source code facilitates an automated and efficient exploration of the decomposition pathways of a molecule by leveraging a combination of reactive molecular dynamics (MD) simulations, graph-theoretic tools, and interactive dashboards. To enhance efficiency, the PES is initially explored using semi-empirical calculations (hereinafter Level1). The stationary points on the PES are then refined using a level of DFT/*ab initio* electronic structure theory (hereinafter Level2) in a subsequent step. Similar to previous studies, Level1 in this work refers to PM7[17] while Level2 is ωB97XD/Def2-TZVPP. The latter is known to perform exceptionally well in predicting barrier heights,[18] and Gaussian09[19] was employed for these calculations. Additionally, thanks to a new algorithm, it is now possible to sample barrierless pathways with our program.[6]

Although AutoMeKin has made extensive use of MD simulations and graph-theoretic tools, the code now includes a new Python library, called *amk_tools*, which enables the parsing, processing, and transformation of the obtained CRNs.[6] Furthermore, new Python scripts are utilized to customize the energy profiles generated by *amk_tools*,[6] and to compare the resulting fragmentation products (R1·, R2·, R3·, and M) with a database of molecules identified in the ISM.

For each system (GLN and GLN-H), AutoMeKin's workflow was iterated 30 times. Each iteration comprised several steps, such as MD simulations, optimization of minima and TSs, and generation of CRNs. For each set of MD simulations, 500 trajectories were performed, with a maximum simulation time of 0.5 ps. The screening settings used in our recent indole decomposition study have been adopted here.[6] These settings are used to filter TSs with very low imaginary frequencies, and to remove redundant structures. Further details on the applied methodology can be found in Ref. 6.

To determine whether the CRNs obtained in our iterative procedure have converged, kinetic Monte Carlo (KMC) simulations were used to solve the chemical master equation.[20] Specifically, the abundances of the products, obtained after exciting 10$^3$ molecules of GLN and GLN-H at 250 kcal/mol, were calculated. RRKM theory was utilized to evaluate the rate coefficients for each state-to-state process:[21]

$$k(E) = \sigma \frac{W^{\ddagger}(E-E_0)}{h\rho(E)} \tag{3}$$

Here, $\sigma$ represents the degeneracy of the reaction coordinate, $W^{\ddagger}(E-E_0)$ is the sum of states of the transition state, $E_0$ is the barrier height, and $\rho(E)$ is the density of states of the reactant. These calculations have been performed using the kinetics module of AutoMeKin, and the sums and density of states are obtained by direct counting, utilizing the Beyer-Swinehart algorithm.[22] Fig. S1 presents the product abundances plotted against iteration number for the decomposition of both GLN and GLN-H; barrierless channels are not included in this analysis. The figure shows the convergence of the results after a few iterations. It is worth noting that the primary products CH$_2$O and C$_2$HNO obtained during the decomposition of GLN have also been experimentally observed in the photodissociation of the molecule.[23]

Although our research is limited to the investigation of the barrierless formation channels of GLN, the full CRNs can be found in Zenodo.[24] In the CRNs, the designations for the minima, transition states, and fragments are MIN, TS, and PROD (or PR), respectively. These designations are accompanied by numerical labels to differentiate between various structures. Additionally, the labels for GLN and GLN-H structures are independent.

### 2.2 Kinetic simulations

The two predominant pathways chosen for the kinetic analysis exhibit two distinct and consecutive dynamical bottlenecks:

i. A free energy bottleneck (TS1) to the barrierless formation of a vibrationally excited species (S$^*$),

$$A + B \rightarrow S^* \tag{4}$$

ii. The second bottleneck (TS2) occurs in the vicinity of a saddle point, where S$^*$ undergoes an elimination reaction:

$$S^* \rightarrow C + D \tag{5}$$

In the low-pressure regime, the canonical unified statistical (CUS)[25, 26] model can be utilized. This model allows us to approximate the rate coefficient for the formation of the C+D products as:

$$k^{\text{CUS}} = \frac{k_1}{k_1+k_2} k_2 \tag{6}$$

In eq. 6, the association rate coefficient, denoted as $k_1$, corresponds to eq. 4. On the other hand, the bimolecular rate coefficient, denoted as $k_2$, represents the passage through TS2, considering TS1 is absent.[25] The calculation of $k_2$ was performed using canonical variational transition state theory (CVT),[27] with A and B as the reacting species. The bimolecular rate coefficients $k_2$ were computed using Pilgrim.[28]

The association rate coefficient $k_1$ was calculated using the formula proposed by Georgievskii and Klippenstein for dipole-dipole interactions,[29] which is expressed as:

$$k_1 = C\mu^{-1/2}(d_A d_B)^{2/3} T^{-1/6} \tag{7}$$

where $\mu$ represents the reduced mass of the two fragments (in amu), $T$ represents the temperature (in K) and $d_A$ and $d_B$ denote the Level2-calculated dipole moments (in D) of the reactants.

The value of $C$ is $1.83 \times 10^{-9}$ with units such that $k_1$ is expressed in cm$^3$ molecule$^{-1}$s$^{-1}$.

The vibrationally excited species S* can also undergo a radiative stabilization process which can potentially compete with reaction 5:

$$S^* \xrightarrow{k_r} S + h\nu \quad (8)$$

The rate coefficient for the radiative stabilization step was calculated using the harmonic approximation:[30]

$$k_r(\text{s}^{-1}) = \sum_{i=1}^{N_m} \sum_{n=0}^{\infty} 1.25 \times 10^{-7} n P_i(n) I_i \nu_i^2 \quad (9)$$

where $N_m$ denotes the total number of vibrational modes, $P_i(n)$ represents the probability of mode $i$ being in level $n$, $I_i$ corresponds to the infrared absorption intensity for the transition from level $n = 0$ to $n = 1$ of mode $i$ (in units of km/mol), and $\nu_i$ is the frequency of the $i$th vibrational mode in cm$^{-1}$. The population distribution can be determined using the following equation:[30]

$$P_i(n, E) = \frac{\rho_{\text{vib}}^{N-1,i}(E - nh\nu_i)}{\rho_{\text{vib}}(E)} \quad (10)$$

In eq. 10, $\rho_{\text{vib}}^{N-1,i}$ refers to the density of states for the molecule with the $i$th mode absent.

## 3 Results

This section outlines the plausible pathways of GLN formation in the ISM, classified into radical-radical and radical-molecule reactions. It is worth noting that one of the pathways discussed below does not fit into either of the two categories mentioned earlier, as it involves reactions between a molecule and a biradical. Despite this, it has been included in the category of reactions between radicals because it was obtained through the study of singlet GLN decomposition, similar to other radical-radical reactions.

The figures displaying the reaction pathways incorporate zero-point energy corrections and the relative energies are referenced with respect to the energy of the starting materials. Primary pathways are shown in the main text, while additional, longer pathways are included in the Electronic Supplementary Information (ESI). The pathways are denoted by the structural formula of the smallest precursor, followed by Roman numerals in the event that multiple pathways of the same type are present.

### 3.1 Radical-radical reactions

**3.1.1 CN + CH$_3$O.** The ISM has been found to contain both cyano (CN)[31] and methoxy (CH$_3$O)[32] radicals. While the hydroxymethyl radical (CH$_2$OH) isomer has not yet been directly detected due to its high chemical reactivity,[33] it has been considered in our analysis since certain pathways involving this isomer lead to GLN in just a few steps.

Fig. 1 shows the only identified pathway for the formation of GLN from CN and CH$_3$O. Additionally, Fig. S2 displays five additional formation pathways (CN-II to CN-VI) from the reactants CN+CH$_2$OH.

The pathways are arranged in decreasing order of relevance for the formation of GLN in the ISM. The CN-I pathway is the sole CN+CH$_3$O (PR44) route discovered in our study for the submerged-barrier formation of GLN. As only the methoxy radical isomer has been identified in the ISM thus far, it is the most pertinent route among those outlined in this section. It involves the association of both radicals to produce methyl cyanate (MIN7), which then transforms into GLN (MIN1) via isomerization. As elaborated below, this isomerization process paves the way for proposing a novel formation mechanism.

The CN-II pathway involves the one-step formation of GLN through the CN+CH$_2$OH (PR24) reaction, which leads to the creation of a C−C bond between the cyano and hydroxymethyl radicals. Moreover, the CN-III, CN-IV, CN-V and CN-VI pathways all start with the same reactants (PR24) and involve the primary formation of isocyanomethanol (MIN8). This intermediate compound is then connected to GLN via one (CN-III), two (CN-IV) or three steps (CN-V and CN-VI).

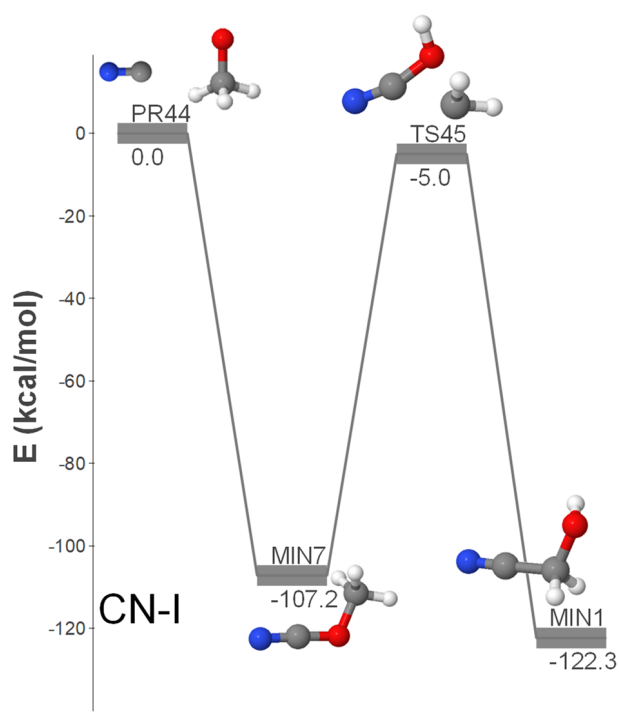

**Fig. 1** DFT-computed energy profile (energy in kcal/mol) for the barrierless formation of GLN from CN+CH$_3$O.

**3.1.2 CH$_2$ + HOCN.** This section delves deeper into a specific part of the larger reaction pathway previously discussed in section 3.1.1. Given the presence of methylene (CH$_2$)[34] and cyanic acid (HOCN)[35] in the ISM, in this section a plausible route for the formation of GLN is explored. The path relies on the transition state connecting methyl cyanate and GLN (TS45), which was examined in the preceding section as part of the CN-I pathway. In particular, our hypothesis suggests that the region around TS45 can be directly reached from the CH$_2$ + HOCN fragments. This suggests that TS45 would serve two distinct reaction pathways, which is a non-statistical feature found in prior studies. [6, 36, 37]

The minimum energy path (MEP) for the isomerization process is shown in Fig. 2. The mechanism bears similarities to roaming,[38] where the methylene and cyanic acid fragments initially separate as if they were undergoing dissociation. However, the two radicals ultimately recombine (frustrating the dissociation) to yield methyl cyanate and GLN. The MEP is quite extensive: on the methyl cyanate side, the methylene fragment initially abstracts a hydrogen atom from the oxygen before subsequently recombining. On the GLN side, the $CH_2$ fragment is inserted into the carbon-oxygen bond. The PES around the saddle point region is rather flat, and the energy of TS45 is just 2 kcal/mol higher than the asymptotic limit of $CH_2$+HOCN. Consequently, depending on their orientation, the arrangement of the fragments can result in the formation of either methyl cyanate or GLN, making both outcomes feasible. Earlier studies reported similar dynamic effects where a single transition state participates in two mechanisms.[6, 36, 37] If such non-statistical behavior also exists for TS45, it would support the feasibility of forming GLN from the $CH_2$ and HOCN fragments. Quasi-classical trajectory simulations could help confirm this possibility.

If the proposed formation pathway from $CH_2$ and HOCN fragments is validated in future research, it is worth noting that the initial excitation energy of GLN would be approximately 120 kcal/mol. This substantial amount of energy increases the likelihood of GLN undergoing further unimolecular reactions, potentially competing with radiative stabilization.

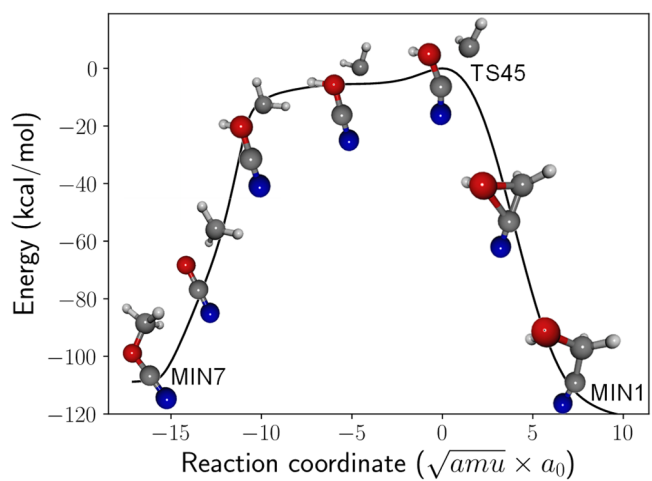

**Fig. 2** DFT-computed minimum energy path (energy in kcal/mol) for roaming-like isomerization between methyl cyanate (MIN7) and glycolonitrile (MIN1) through TS45.

**3.1.3 OH + $CH_2CN$.** Hydroxyl[39] and cyanomethyl ($CH_2CN$)[40] are two radicals that have already been confirmed to be present in the ISM. Due to their low energy relative to GLN (~79 kcal/mol) and ability to combine and form GLN in a single step (see the upper panel of Fig. 3; pathway OH-I), OH and $CH_2CN$ (PR25) are ideal candidates to serve as precursors. Since the total energy of the two radicals is low, the formation of GLN would result in a limited number of isomerizations or dissociations, as discussed below. Pathway OH-II (shown in Fig. S3) represents an alternative formation route from OH+$CH_2NC$ (PR25). However, it is worth noting that the $CH_2NC$ radical has not been observed in the ISM yet.

Based on the reasons outlined in this section, it can be concluded that the OH+$CH_2NC$ reaction is the primary radical-radical pathway for the formation of GLN identified in this study. As a result, the kinetic simulations described in Section 3.3 will be focussed on this particular radical-radical mechanism.

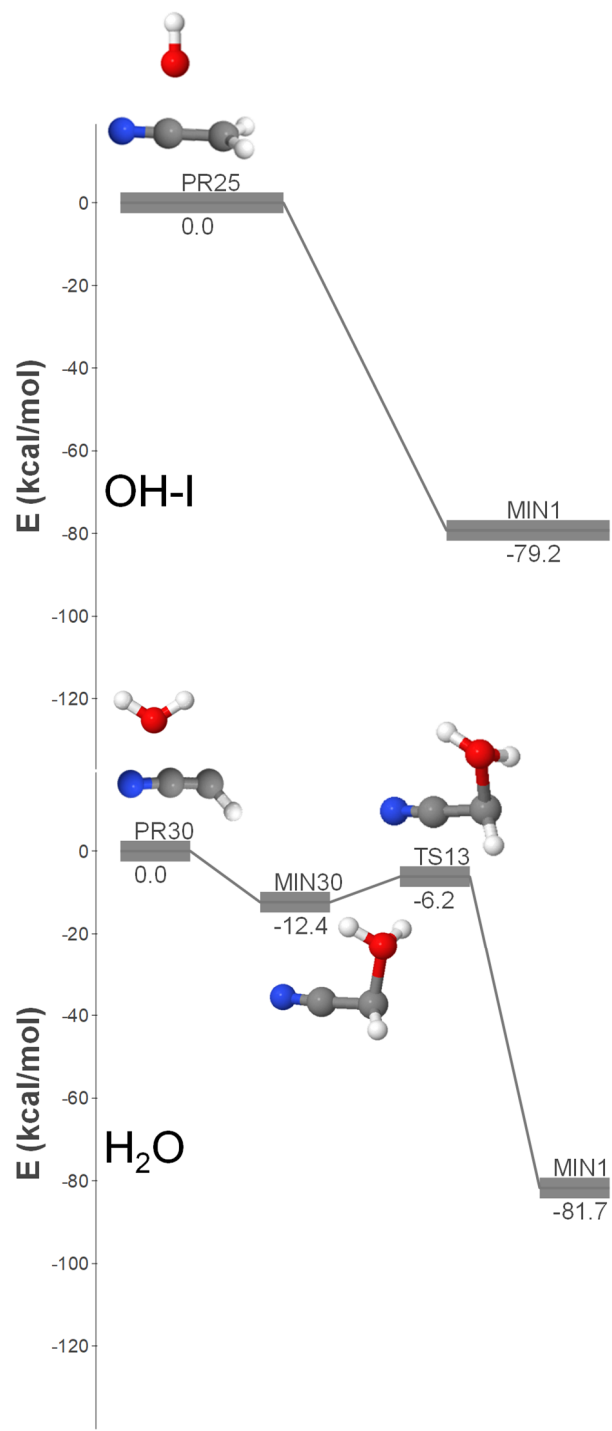

**Fig. 3** DFT-computed energy profiles (energy in kcal/mol) for the barrierless formation of GLN from OH+CH$_2$CN (OH-I pathway) and H$_2$O+HCCN (H$_2$O pathway).

#### 3.1.4 H$_2$O + HCCN.

Water[41] and cyano methylene (HCCN)[42] are another set of feasible precursors, both of which have been detected in the ISM. As previously mentioned, this pathway does not pertain to the category of reactions between two radicals, but has been included in this section, nonetheless. HCCN exhibits a strong biradical nature, with the ground state (triplet) being approximately 10 kcal/mol lower in energy than the singlet state.

The lower panel of Fig. 3 (pathway H$_2$O) shows that the association of H$_2$O and HCCN (PR30) results in the formation of a highly unstable intermediate (MIN30), which can undergo subsequent isomerization via hydrogen migration to yield GLN. The high energy of MIN30 significantly reduces the radiative stabilization rate compared to pathway OH-I. Moreover, the formation of GLN requires an additional step, which diminishes the kinetic significance of this pathway relative to that of OH-I. Put simply, the formation rate coefficient for this pathway would be reduced.

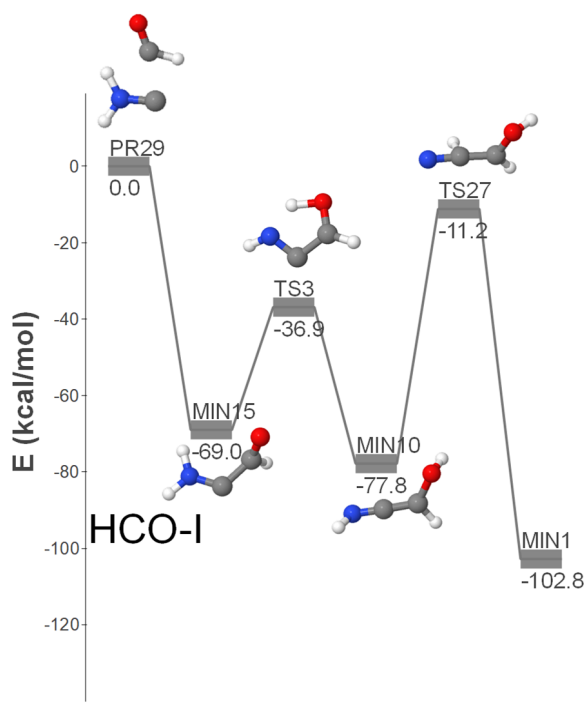

**Fig. 4** DFT-computed energy profile (energy in kcal/mol) for the barrierless formation of GLN from HCO and H$_2$NC (HCO-I pathway).

#### 3.1.5 HCO + CH$_2$N.

A total of five isomers of the reactants have been found in this study to connect via submerged barriers with GLN. These include the formyl radical (HCO; detected in the ISM[43]), the isoformyl radical (HOC; only its cation has been detected in the ISM[44, 45]), the methylene amidogen radical (H$_2$CN; detected in the ISM[46]), the aminocarbyne radical (H$_2$NC; detected in the ISM[47]), and the formimidoyl radical (HNCH; not detected in the ISM).

Fig. 4 shows the HCO-I pathway, the only three-step route that incorporates two reactants detected in the ISM: namely formyl and aminocarbyne radicals (PR29). An alternative (longer) pathway, HCO-VII, shown in Fig. S5, connects MIN10 with GLN via an additional step. Fig. S5 shows additional GLN formation pathways that begin with isomers of the reactants that have not yet been detected in the ISM. These pathways could become relevant in the event of new detections.

Pathway HCO-I is initiated with the formation of MIN15, followed by two isomerization steps that involve hydrogen migrations over submerged barriers and lead to the formation of GLN. Due to the relatively high energy of the reactants and the involvement of three elementary steps, this pathway is considered less significant than OH-I.

### 3.2 Radical-molecule reactions

This study has identified only two radical-molecule pathways that can lead to the formation of GLN in a single step. One of these pathways is the reaction between the hydroxyl radical and methyl cyanide:

$$\text{OH} + \text{CH}_3\text{CN} \rightarrow \text{GLN} + \text{H} \qquad (11)$$

The second mechanism involves the reaction between the hydroxymethyl radical and hydrogen cyanide (HCN):

$$\text{CH}_2\text{OH} + \text{HCN} \rightarrow \text{GLN} + \text{H} \qquad (12)$$

The transition states for these two reactions are TS177 and TS201 of the GLN-H CRN, respectively. Despite the detection of HCN[48] and CH$_3$CN[49] in the ISM, the barrier heights for the radical-molecule mechanisms described above are significant, with 50.3 and 60.0 kcal/mol with respect to reactants, for reactions (11) and (12), respectively.

We have discovered another potentially interesting reaction in the ISM:

$$\text{OH} + \text{CH}_3\text{CN} \rightarrow \text{H}_2\text{O} + \text{CH}_2\text{CN} \qquad (13)$$

However, the transition state corresponding to this process (TS34) has an energy higher than that of the reactants, which makes it unfavorable. The measured rate coefficient for this reaction within the temperature range of 256-388 K is of the order of $10^{-14}$ cm$^3$ molecule$^{-1}$ s$^{-1}$.[50] We have determined the rate coefficient for reaction (13) at 10 K using the canonical unified statistical theory[25, 26] under the low-pressure limit. Within these pressure conditions, the tunneling effect was taken into account using the Pilgrim program[28] with the small-curvature approximation, considering tunneling energies above the energy of the reactants. The resulting rate coefficient is significantly lower than $10^{-14}$ cm$^3$ molecule$^{-1}$ s$^{-1}$.

However, the addition reaction between OH and CH$_3$CN leading to CH$_3$C(OH)N, competes with reaction (13) under low-pressure conditions at 10 K, with a computed rate coefficient of approximately $5\times10^{-14}$ cm$^3$ molecule$^{-1}$ s$^{-1}$. As the temperature exceeds 200 K, the addition reaction becomes pressure-independent, and our calculated values align well with experimental data within the temperature range of 256 to 388 K.[51]

We have also investigated pathways that involve multiple elementary steps and have identified the following reaction, whose energy profile is depicted in Fig. 5:

$$OH + CH_2CNH \rightarrow GLN + H \tag{14}$$

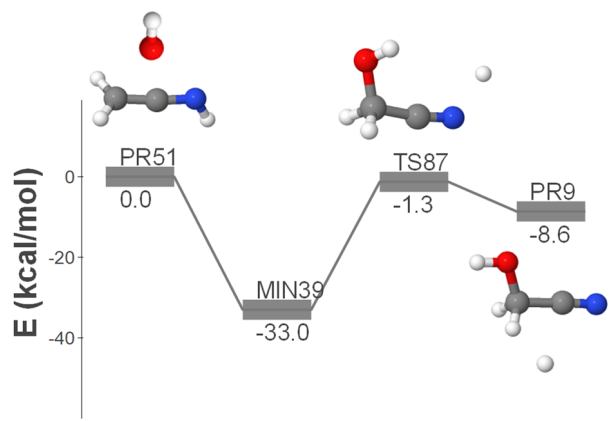

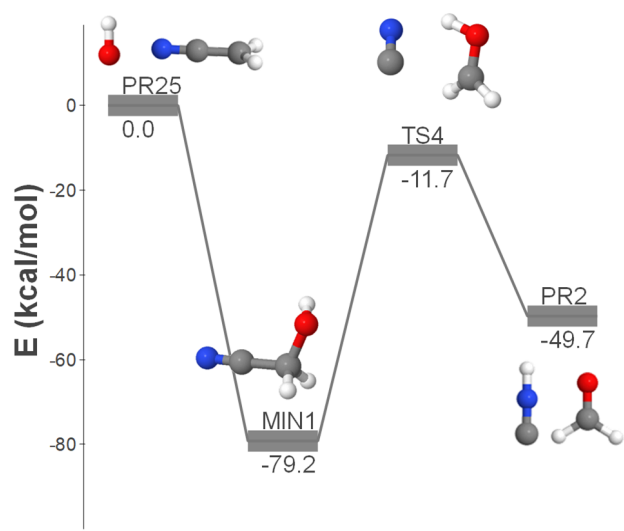

**Fig. 6** DFT-computed energy profile (energy in kcal/mol) for the formation of HNC+CH$_2$O from OH+CH$_2$CN.

**Fig. 5** DFT-computed energy profile (energy in kcal/mol) for the barrierless formation of GLN+H (PR9) from OH and CH$_2$CNH (R-M pathway). Note that the numerical labels of the structures in this figure are not related to those shown in the other figures.

The reactants of reaction 14 are the hydroxyl radical and ketenimine (CH$_2$CNH), which was first detected in the ISM in 2006.[52] This reaction proceeds through only two elementary steps, making it an ideal candidate for a gas phase process for the formation of GLN. The initial association results in the formation of MIN39, which ultimately loses a hydrogen atom, leading to the formation of GLN and atomic hydrogen.

Fig. S6 shows an alternative, longer route, which involves the aminocarbyne radical and formaldehyde as reactants. The presence of the latter was first detected in the ISM in 1969.[53] This reaction mechanism proceeds via the highly unstable intermediate MIN82, which rearranges via H-migration to form MIN39. This pathway shares the final dissociation step with the one shown in Fig. 5.

### 3.3 Kinetic simulations

Among the radical-radical reactions, pathway OH-I, which involves the combination of OH and CH$_2$CN to form GLN in a single step, is the most promising and efficient route for the formation of GLN. However, as shown in Fig. 6, GLN can potentially dissociate into hydrogen isocyanide (HNC) and formaldehyde (CH$_2$O) by surmounting a barrier of 67.6 kcal/mol, making the formation of these two products possible.

The initially formed glycolonitrile molecules exhibit a unimolecular rate of HNC+CH$_2$O formation (~$10^6$ s$^{-1}$), which is four orders of magnitude higher than their rate of radiative stabilization (~$10^2$ s$^{-1}$). This suggests that the formation of GLN through this pathway is not efficient. Nevertheless, due to the submerged nature of TS4, the following process becomes an efficient pathway for the formation of hydrogen isocyanide and formaldehyde:

$$OH + CH_2CN \rightarrow HNC + CH_2O \tag{15}$$

Fig. 7 shows that the low-pressure limit rate coefficient for this process is approximately $10^{-9}$ cm$^3$ molecule$^{-1}$ s$^{-1}$ within the temperature range of 10-150 K.

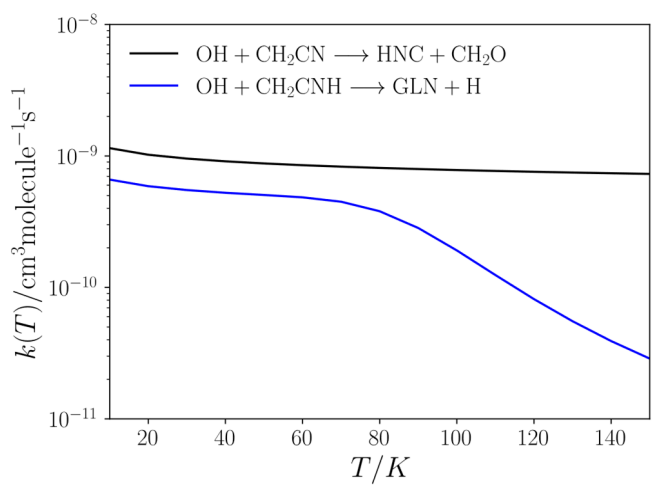

**Fig. 7** Low-pressure rate coefficients for reactions 14 (blue) and 15 (black) calculated using the CUS model.

The fact that, in eq. (15), the resulting product is HNC instead of HCN is significant and could potentially help to explain the observed overabundance of HNC in the ISM, as compared to its expected equilibrium value.[54, 55] Various previous studies have attempted to account for this discrepancy by proposing possible photolytic sources of HCN and HNC in the ISM. These studies have reported measured or calculated HNC/HCN branching ratios in the range of 0.1-0.3.[56-59]

The discovery of the efficient formation of HNC from GLN enables us to propose all barrierless radical-radical reactions of the present study as viable sources of HNC in the interstellar medium (ISM), albeit with slightly lower efficiency compared to reaction 15.

In this section we have also examined the rate coefficients for the radical-molecule process displayed in Fig. 5 (reaction 14). This pathway exhibits two elementary steps that mimic reaction 15. Fig. 7 presents the rate coefficients for this reaction in the low-pressure regime, covering the temperature range of 10-150 K. In reaction 15, the formation of GLN is identified as the rate-determining step throughout the entire temperature range. However, in reaction 14, the rate-determining step is the formation of GLN-H, but only until approximately 90 K. At this point the contribution of $k_2$ to the overall formation rate coefficient becomes significant (see Table S2). This observation explains the noticeable decline in the rate coefficient of reaction 14 (see Fig. 7).

## 4 Conclusions

In this study, we have employed automated reaction discovery methods along with CUS and CVT kinetic calculations to investigate the possible formation pathways of glycolonitrile. Our findings can be summarized as follows:

1. Our findings indicate that the gas-phase reactions involving hydroxyl radical with cyanomethyl and ketenimine are the most promising pathways for the formation of glycolonitrile.
2. In the reaction between hydroxyl and cyanomethyl radicals, glycolonitrile undergoes further decomposition, resulting in the formation of hydrogen isocyanide and formaldehyde. While this finding rules out this pathway as a viable alternative for glycolonitrile formation, it reveals an intriguing gas-phase reaction that significantly contributes to the efficient production of interstellar HNC.
3. The discovery of a highly efficient HNC elimination pathway from GLN expands the potential for any radical-radical mechanism proposed in our study to serve as a viable formation route for HNC in the interstellar medium (ISM).
4. The only viable gas-phase pathway for glycolonitrile formation identified in this study is the reaction between hydroxyl with ketenimine, which also generates atomic hydrogen.
5. The kinetic calculations performed in this study reveal the efficiency of both hydrogen isocyanide and glycolonitrile formation under the harsh conditions of the interstellar medium.

## Conflicts of interest

There are no conflicts of interest to declare.

## Acknowledgements

This work was supported by the Spanish Ministry of Science and Innovation (MICINN) through the CHEMLIFE project (Ref. PID2020-113936GB-I00). It was also partially supported by the Consellería de Cultura, Educación e Ordenación Universitaria (Centro singular de investigación de Galicia acreditación 2019-2022, ED431G 2019/03 and Grupo de referencia competitiva ED431C 2021/40) and the European Regional Development Fund (ERDF), and MICINN through Grant #PID2019-107307RB-I00. ALS thanks Xunta de Galicia for financial support through a predoctoral grant.